\documentclass[%
 reprint,
 amsmath,amssymb,
]{revtex4-1}

\usepackage{graphicx}% Include figure files
\usepackage{dcolumn}% Align table columns on decimal point
\usepackage{bm}% bold math
\usepackage[colorlinks,linkcolor=blue,anchorcolor=blue,citecolor=blue]{hyperref}
\begin{document}

\preprint{APS/123-QED}

\title{Bias-free source-independent quantum random number generator
}% Force line breaks with \\

\author{Ziyong Zheng$^{1}$}
% \thanks{Correspondence: zhangyc@bupt.edu.cn.}%
\author{Yichen Zhang$^{1}$}%
\email{Correspondence: zhangyc@bupt.edu.cn.}
\author{Min Huang$^{2}$}%
\author{Ziyang Chen$^{2}$ }%
\author{Song Yu$^{1}$}%
\author{Hong Guo$^{2}$}%
\email{Correspondence: hongguo@pku.edu.cn.}
\affiliation {%
 $^{1}$ State Key Laboratory of Information Photonics and Optical Communications, Beijing University of Posts and Telecommunications, Beijing, 100876, China \\
 $^{2}$ State Key Laboratory of Advanced Optical Communication Systems and Networks, Department of Electronics, and Center for Quantum Information Technology, Peking University, Beijing 100871, China%\textbackslash\textbackslash
}%

%\affiliation{ Second institution and/or address\\ This line break forced% with \\}%
%\affiliation{Third institution, the second for Charlie Author}%
%\affiliation{%Authors' institution and/or address\\This line break forced with \textbackslash\textbackslash}%


\date{\today}% It is always \today, today,

\begin{abstract}
A bias-free source-independent quantum random number generator scheme based on the measurement of vacuum fluctuation is proposed to realize the effective elimination of system bias and common mode noise introduced by the local oscillator. Optimal parameter settings are derived to avoid the system recording two canonically conjugate quadratures simultaneously in each measurement. In particular, it provides a new approach to investigate the performance difference between measuring two quadratures of equal and unequal intensity. It is experimentally demonstrated that the system supports 4.2 Gbps bias-free source-independent random number generation, where its common mode rejection ratio reaches 61.17 dB. Furthermore, the scheme offers an all-optical method facilitating the integration of source-independent quantum random number generators into compact chips.
\end{abstract}


\maketitle

%\tableofcontents

\section{\label{set1}Introduction}

Quantum random number generator (QRNG), which exploits the intrinsic probabilistic quantum processes to generate random numbers, is theoretically considered to be the most possible way to obtain true random numbers \cite{Ma2016Quantum,Herrero2017quantum,Bera2017Randomness}. However, the practical imperfect devices that introduce noise into the output signals will inevitably compromise the security of QRNG systems. Particularly, the quantum source, where the true randomness originates, acts as the most complicated component in the QRNG system and its fine characterization is usually absent. The security loophole that the quantum source might be prepared or manipulated by the malicious eavesdropper is difficult for the user Alice to perceive in practical situations. To fill the gap, the source-independent quantum random number generator (SI-QRNG) protocols \cite{marangon2017source,avesani2018source}, which release the assumptions on the input state by trusting the measurement devices can fully characterize all the measured signals, enable the generation of unpredictable random numbers with untrustworthy source.
	
	Differing from the discrete-variable SI-QRNG protocol \cite{cao2016source}, the continuous-variable SI-QRNG protocols exploit the high-dimensional nature of the quantum source and have been proposed and demonstrated to be able to achieve faster random number generation speed up to Gbps. Up to now, vacuum fluctuation \cite{Gabriel2010A,Haw2015Maximization,Santamato2017An,gehring20188} and phase noise \cite{Xu2012Ultrafast,Abellan2014Ultra,Yang2016A,Huang2020Aphase} are two main continuous-variable quantum sources for random number generation, where vacuum fluctuation has become a research focus recently because the model of SI-QRNG based on measuring vacuum fluctuation is relatively simple and it supports the implementation of a stable and integrated SI-QRNG system that is insensitive to the detection efficiency. As a promising quantum random source, vacuum fluctuation has already been widely exploited in the analysis and implementation of continuous-variable SI-QRNG protocols.

	%The vacuum fluctuation is so weak to be directly measured that a local oscillator (LO) is required to amplify it through the interference between LO and vacuum state, where the available intensity of LO in the SI-QRNG system will inevitably limited by the practical asymmetrical devices with asymmetrical beam splitter and photodiodes. More importantly,
	Existing SI-QRNGs assume a constant intensity of local oscillator (LO) \cite {xu2019high,zhang2020finite}, which is not consistent with the facts and detailed analysis of eliminating the LO fluctuation in the SI-QRNG scenario is still absent. The residual common mode noise introduced by the fluctuated LO in the biased system will inevitably lead to the overestimation of true randomness, which will definitely compromise the security of generated random numbers. So far, technologies, i.e., frequency mixing \cite{Shen2010Practical,Symul2011Real,guo2018enhancing,guo2019parallel}, alternating-current (AC) coupling \cite{ZHENG20186Gbps} and optical interfering \cite{Zheng2019Experimental,huang2019integrated}, have been tried to eliminate the system bias together with the common mode noise introduced by the LO. However, the frequency mixing technology processes the detected signal after amplification, which works under the conditions of unsaturated measurement and intuitively, it can't do anything to avoid saturation in the trans-impedance amplifier. Besides, the contribution of AC coupling technique on eliminating common mode noise except for the DC component is limited, which still affects the security of the system due to the remaining common mode noise.
	
	Integrated quantum photonics offers an approach of integrating quantum optical components into monolithic structures \cite{wang2019integrated,zhang2019integrated}, and recently the research of integrated QRNG systems has become a  hotspot\cite{abellan2018integrated,raffaelli2018generation,rude2018interferometric,roger2019real,imran2020quantum}. The optical interfering technology based on Mach Zehnder interferometer (MZI) structure offers an all-optical bias elimination technology, which supports the realization of chip integration based on the existing photonic technologies and its feasibility has been verified in practical QRNGs based on measuring vacuum fluctuation. Counterintuitively, we will prove in Sec.2 that each measured signal will simultaneously contain two canonically conjugate quadratures, i.e., X quadrature and P quadrature, by directly applying the existing MZI structure, which violates the requirements of implementing the SI-QRNG protocol. A necessary optimization of the system is required to realize measuring only a single quadrature in each measurement, which extensively enables the realization of SI-QRNG system with three different routines and provides a new method to investigate the influence of symmetrical and asymmetrical measurement of quantum quadratures.
	
	In this work, we focus on solving the bias problem introduced by the practical unbalanced devices applied in a SI-QRNG system. Considering in the untrustworthy source scenario, here we put forward and demonstrate a bias-free scheme where quantum origin of vacuum fluctuation can be exploited for the generation of source-independent quantum random numbers. The scheme utilizes only one MZI structure to realize the effective removal of system bias and well elimination of common mode noise introduced by the LO. What's more, it explores a robust and bias-free SI-QRNG structure suitable for system integration based on the existing technologies, which makes SI-QRNGs low cost and high practical security in the future. Simultaneously, the system parameters are optimized to seek for measuring only a single quadrature in each measurement, which can be further exploited to realize the SI-QRNG system under three different routines. In particular, the optimized system provides a new approach to investigate the performance difference between measuring symmetrical and asymmetrical quadratures by measuring two quadratures of equal and unequal intensity. Combining with the theory of the extremality of Gaussian states, we experimentally implement the system to reach a random number generation speed of 4.2 Gbps. Besides, the final random numbers have passed all the NIST-STS test items.

\section{\label{set2}Architecture design and theoretical model}
\begin{figure*}[t]
	\centering
	\includegraphics[width = 15 cm]{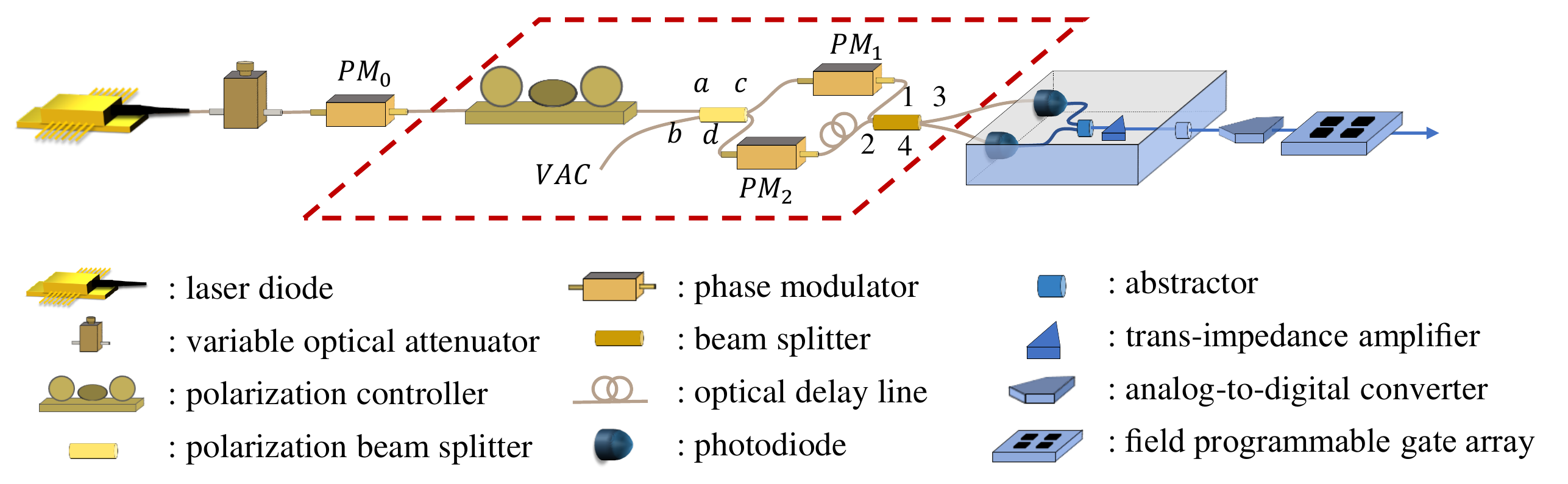}
	\caption{\label{figure_system} The architecture of our proposed SI-QRNG. The structure within the red dotted frame is the novel structure used to achieve bias-free SI-QRNG system. The phase difference of these two arms should maintain a stable value of $\Delta \varphi $.}
	%  用于交叉引用
\end{figure*}

The architecture of the proposed bias-free SI-QRNG setup is depicted in Fig.\ref{figure_system}. The continuous-wave linearly polarized light beam emitted by the 1550 nm fiber-coupled DFB laser diode (NKT, Basik E15) will be modulated by the phase modulator and rotated by the polarization controller (PC) with negligible bend loss. The output light beam will then interfere with the vacuum state introduced from the physically blocked port of the $2 \times 2$ polarization beam splitter (PBS), where the interfered signal will be split into two orthogonal polarization directions and be modulated by the phase modulators separately in the upper and lower arms. A compensation phase $\Delta \varphi $ will be loaded on the phase modulator in the upper arm in our system to eliminate the bias introduced by the asymmetric devices. The corresponding power splitting ratio will be changed by adjusting the polarization direction of PC.	A further beam splitter (BS) supports the interference of signals from these two arms and the two outputs will be directly coupled into a DC coupled homodyne detector (INSIGHT, BPD-1). To obtain high-speed digital random numbers, the analog-to-digital converter (ADC, TI, ADS5400) will be applied to transform the analog signals into digital bits for the convenience of further randomness extraction, which will be operated at the field programmable gate array platform.

	Theoretical model of the setup is established as follows. As shown in Fig.\ref{figure_system}, suppose the electric fields of LO and the measured vacuum state are ${E_L}(t) = {E_L} + \delta {X_L}(t) + i{P_L}(t)$ and ${E_s}(t) = {E_s} + \delta {X_s}(t) + i{P_s}(t)$, where ${E_L}$ and ${E_s}$ are time-independent terms, and $\delta {X_{L(s)}}(t)$ and $\delta {P_{L(s)}}(t)$ are time-dependent items that describe the changes of X and P quadratures of LO (vacuum state) field. Generally, a phase modulator will be exploited to shift the phase of LO, which results to a phase difference of ${\varphi}$ between LO and vacuum state. The electric fields at port c and d will be
	\begin{equation}
	\label{Ecd}
	\begin{split}
	\left[ {\begin{array}{*{20}{c}}
{{E_c}(t)}\\
{{E_d}(t)}
\end{array}} \right] = \left[ {\begin{array}{*{20}{c}}
{\sqrt {{t_{ac}}} }&{\sqrt {{r_{bc}}} }\\
{\sqrt {{r_{ad}}} }&{ - \sqrt {{t_{bd}}} }
\end{array}} \right]\left[ {\begin{array}{*{20}{c}}
{{\eta _{P{M_0}}}{E_L}(t){e^{i\varphi }}}\\
{{E_s}(t)}
\end{array}} \right],
	\end{split}
\end{equation}
where ${\eta _{P{M_0}}}$ is the insertion loss of ${P{M_0}}$, ${t_{ac}}$, ${{t_{bd}}}$, ${r_{ad}}$, and ${r_{bc}}$ are the transmission and reflection coefficients of PBS. Assuming the insertion loss of ${P{M_1}}$ and ${P{M_2}}$ are ${\eta _{P{M_1}}}$ and ${\eta _{P{M_2}}}$ respectively, the electric fields at port 3 and 4 will be expressed as	
	\begin{equation}
	\label{E34}
	\begin{split}
	\left[ {\begin{array}{*{20}{c}}
{{E_3}(t)}\\
{{E_4}(t)}
\end{array}} \right] = \left[ {\begin{array}{*{20}{c}}
{\sqrt {{t_{13}}} }&{\sqrt {{r_{23}}} }\\
{\sqrt {{r_{14}}} }&{ - \sqrt {{t_{24}}} }
\end{array}} \right]\left[ {\begin{array}{*{20}{c}}
{{\eta _{P{M_1}}}{E_c}(t){e^{i\Delta \varphi }}}\\
{{\eta _{P{M_2}}}{E_d}(t)}
\end{array}} \right],
	\end{split}
\end{equation}
where ${t_{13}}$, ${t_{24}}$, ${r_{14}}$ and ${{r_{23}}}$ are the transmission and reflection coefficients of BS.

The homodyne detector transforms the optical signal into electric current and then turns the electric current into voltage signal by using its trans-impedance amplifier. Provided the gains of two photodiodes are ${g_{P{D_1}}}$ and ${g_{P{D_2}}}$ separately, the final output $v$ will be derived as
\begin{widetext}
	\begin{equation}
	\label{EV1}
	\begin{split}
v &= {g_{P{D_1}}}E_3^2(t) - {g_{P{D_2}}}E_4^2(t)\\
 &= A{E_c}(t)E_c^*(t) + B{E_d}(t)E_d^*(t)
 + C[{e^{i\Delta \varphi }}{E_c}(t)E_d^*(t) + {e^{{\rm{ - }}i\Delta \varphi }}{E_d}(t)E_c^*(t)],
	\end{split}
\end{equation}
\end{widetext}
where $A = ({g_{P{D_1}}}{t_{{\rm{13}}}}{\rm{ - }}{g_{P{D_2}}}{r_{{\rm{14}}}})\eta _{P{M_1}}^2$, $B = ({g_{P{D_1}}}{r_{{\rm{23}}}}{\rm{ - }}{g_{P{D_2}}}{t_{24}})\eta _{P{M_2}}^2$, $C = ({g_{P{D_1}}}\sqrt {{t_{{\rm{13}}}}{r_{{\rm{23}}}}}  + {g_{P{D_2}}}\sqrt {{r_{{\rm{14}}}}{t_{24}}} ){\eta _{P{M_1}}}{\eta _{P{M_2}}}$. Notably, the calculations of ${E_c}(t)E_c^*(t)$, ${E_c}(t)E_d^*(t)$, ${E_d}(t)E_c^*(t)$ and ${E_d}(t)E_d^*(t)$ are realized by assuming the infinitesimals $\delta X_L(t)\delta X_s(t)$, $\delta X_L(t)\delta P_s(t)$, $\delta P_L(t)\delta X_s(t)$, $\delta P_L(t)\delta P_s(t)$, $\delta X_L^2(t)$, $\delta P_L^2(t)$, $\delta X_s^2(t)$ and $\delta P_s^2(t)$ are approximately equal to 0. Simultaneously, the value of $E_s$ is treated as 0 due to the reason that we consider an untrusted source in a quantum state with zero mean. In this case, the final output of vacuum fluctuation, which is associated with $\varphi$, can be thus obtained as
\begin{widetext}
	\begin{equation}
	\label{EV2}
	\begin{split}
v &= [A{t_{ac}} + B{r_{ad}} + 2C\sqrt {{t_{ac}}{r_{ad}}} \cos (\Delta \varphi )](\eta _{P{M_0}}^2E_L^2 + 2{\eta _{P{M_0}}}{E_L}\delta {X_L})\\
 &+ 2{\eta _{P{M_0}}}{E_L}[A\sqrt {{t_{ac}}{r_{bc}}}  - B\sqrt {{r_{ad}}{t_{bd}}}  + C(\sqrt {{r_{bc}}{r_{ad}}}  - \sqrt {{t_{ac}}{t_{bd}}} )\cos \Delta \varphi ][\delta {X_s}(t)\cos \varphi  + \delta {P_s}(t)\sin \varphi ]\\
 &+ 2C{\eta _{P{M_0}}}{E_L}(\sqrt {{r_{bc}}{r_{ad}}}  + \sqrt {{t_{ac}}{t_{bd}}} )\sin \Delta \varphi [\delta {X_s}(t)\sin \varphi  + \delta {P_s}(t)\cos \varphi ].
	\end{split}
\end{equation}
\end{widetext}

Besides, the system bias and the common mode noise introduced by the LO can be well eliminated by setting the compensation phase $\Delta \varphi$ as
\begin{equation}
	\label{Deltavarphi}
	\begin{split}
\Delta \varphi  = \arccos \left( {  \frac{-{A{\xi ^{1/2}} - B{\xi ^{ - 1/2}}}}{{2C}}} \right),
	\end{split}
\end{equation}
where $\xi={t_{ac}/{r_{ad}}}$ indicates the power splitting ratio of mode a in the PBS. What's more, the reflection and transmission coefficients of mode b in the PBS, i.e., $r_{bc}$ and $t_{bd}$, will depend on the polarization of the incoming signal and they are not easy for Alice to predict their values in advance. The rotation of the polarization will compromise of the evaluated extractable randomness whilst it can be easily noticed if Alice monitors the statistical variances.

It is counterintuitive that each measured signal will contain two quadratures simultaneously by directly applying the MZI structure without system optimization, as shown in Eq.\ref{EV2}, which is different from the general case where only the X quadrature will be measured when the phase $\varphi$ of $PM_0$ is set as 0 and the P quadrature will be measured when $\varphi=\pi/2$. To ensure the system measures a single quadrature in each measurement, a necessary optimization is required by adjusting the system according to the derived system parameters, which extensively offers the system the ability to work in three different routines.

The output in the first routine can be expressed as
$v=2{\eta _{P{M_0}}}{E_L}[ A\sqrt {{t_{ac}}{r_{bc}}}  - B\sqrt {{r_{ad}}{t_{bd}}}  - C( {\sqrt {{r_{bc}}{r_{ad}}}  - \sqrt {{t_{ac}}{t_{bd}}} } ) ][\delta {X_s}(t)\cos \varphi + \delta {P_s}(t)\sin \varphi ]$, where the corresponding compensation phase $\Delta \varphi$ is set as $\pi$ and $\xi$ should equal to $ (2{C^2} - AB + 2C\sqrt {{C^2} - AB} )/{A^2}$. It supports the measurement of X and P quadratures when $\varphi$ is calibrated as 0 and $\pi/2$.
Differing from the first routine, the actual measured quadrature in the second routine will not be the quadrature to be measured, but its conjugate quadrature, where the output can be given by $v = 2{\eta _{P{M_0}}}C{E_L}(\sqrt {{r_{bc}}{r_{ad}}}  + \sqrt {{t_{ac}}{t_{bd}}} )\sin \Delta \varphi [\delta {X_s}(t)\sin \varphi  + \delta {P_s}(t)\cos \varphi ]$. The X quadrature will be measured when $\varphi=\pi/2$ and the P quadrature will be measured when $\varphi=0$. It should be noticed that the second routine will establish on the premise of $\xi  = B/A$ and $\Delta \varphi= \arccos ( - \sqrt {AB/{C^2}})$. Notably, the intensity of measured X and P quadratures will be equal both in the first two routines. The third routine acts as the combination of previous two routines and will be able to realize switching the measured quadratures by simultaneously adjusting the compensation phase $\Delta \varphi$ and power splitting ratio $\xi$, where the intensity of measured X and P quadratures will be unequal and the switch of measured quadratures no longer depends on the $P{M_0}$. This provides a new approach to investigate the performance of measuring two quadratures of unequal intensity in a homodyne detection system, which is not easy to realize by applying the existing SI-QRNG structures. What's more, by comparing with the measured results of the first routine and second routine, the performance difference between measuring two quadratures of equal and unequal intensity will be obtained. When the phase of LO is set as $\pi/2$, the system will record X quadrature when $\Delta \varphi $ and $\xi$ are set according to the second routine and P quadrature will be measured when $\Delta \varphi $ and $\xi$ are set according to the first routine.

To qualify the extractable randomness of our proposed scheme, here we refer to Ref. \cite{xu2019high} and exploit the theory of the extremality of Gaussian states to analyze the feasibility of the proposed scheme. The covariance matrix ($CM$) of these two measured quadratures $X$ and $P$ of measured quantum state ${\rho _A}$, which acts as a tool to estimate the bound of extractable randomness, can be written as ${\gamma _A} = \left( {\begin{array}{*{20}{c}}
{{V_x}}&c\\
c&{{V_p}}
\end{array}} \right)$,
\iffalse
\begin{equation}
	\label{gammaA}
	\begin{split}
		{\gamma _A} = \left( {\begin{array}{*{20}{c}}{{V_{x}}}&{c}\\
				{c}&{{V_{p}}}\end{array}} \right),
	\end{split}
\end{equation}
\fi
where ${{V_{x}}}$ and ${{V_{p}}}$ are the variances of $X$ and $P$ quadratures and ${c}$ is the co-variance between $X$ and $P$ quadratures. Notably, the values of ${{V_{x}}}$ and ${{V_{p}}}$ will be equal in the first two routines and unequal in the third routine. Similar to the security analysis in the homodyne-based SI-QRNG \cite {xu2019high}, when combining with the theory of the extremality of Gaussian states, the lower bound of the extractable randomness of per measurement conditioned on the existence of eavesdropper can be derived as
\begin{equation}
	\label{Rdis1}
	\begin{split}
		{R_{{\rm{dis}}}}\left( {{a_{xi}}|E} \right) \ge H\left( {{a_{xi}}} \right) - S\left( {\rho _A^G} \right),
	\end{split}
\end{equation}
where $H\left( {{a_{xi}}} \right)$ is the Shannon entropy of quadrature $X$'s discrete variable ${a_{xi}}$, ${\rho _A^G}$ is a Gaussian state with the same $CM$ as ${\rho _A}$ and the above relationship will still hold when switching X quadrature and P quadrature. The Holevo's bound of $\rho _A^G$ can be calculated as
$S\left( {\rho _A^G} \right) \le [(\lambda  + 1)/2]{\log _2}[(\lambda  + 1)/2] - [(\lambda  - 1)/2]{\log _2}[(\lambda  - 1)/2]$, where $\lambda = \sqrt {\det \left({{\gamma _A}} \right)}  = \sqrt {{V_{x}}{V_{p}} - {c^2}}$.

Due to the finite sampling resolution compromising the characterization of the exact values of $\lambda $ and $c$, necessary treatments of setting $c=0$, ${V_{x}}= \overline {{V_x}}$ and ${V_{p}}= \overline {{V_p}}$ help to obtain a upper bound of $\lambda $ and finally a lower bound of ${R_{{\rm{dis}}}}\left( {{a_{xi}}|E} \right)$. The values of $\overline {{V_x}}$ and $\overline {{V_p}}$ can be calculated by treating ${a_i}$ as ${a_i} - 0.5\Delta $ when ${a_i} \le 0$ or ${a_i} + 0.5\Delta $ when ${a_i} > 0$, where $\Delta $ is the digitization interval of ADC.

\begin{figure}[t]
	\centering
	\includegraphics[width = 9
 cm]{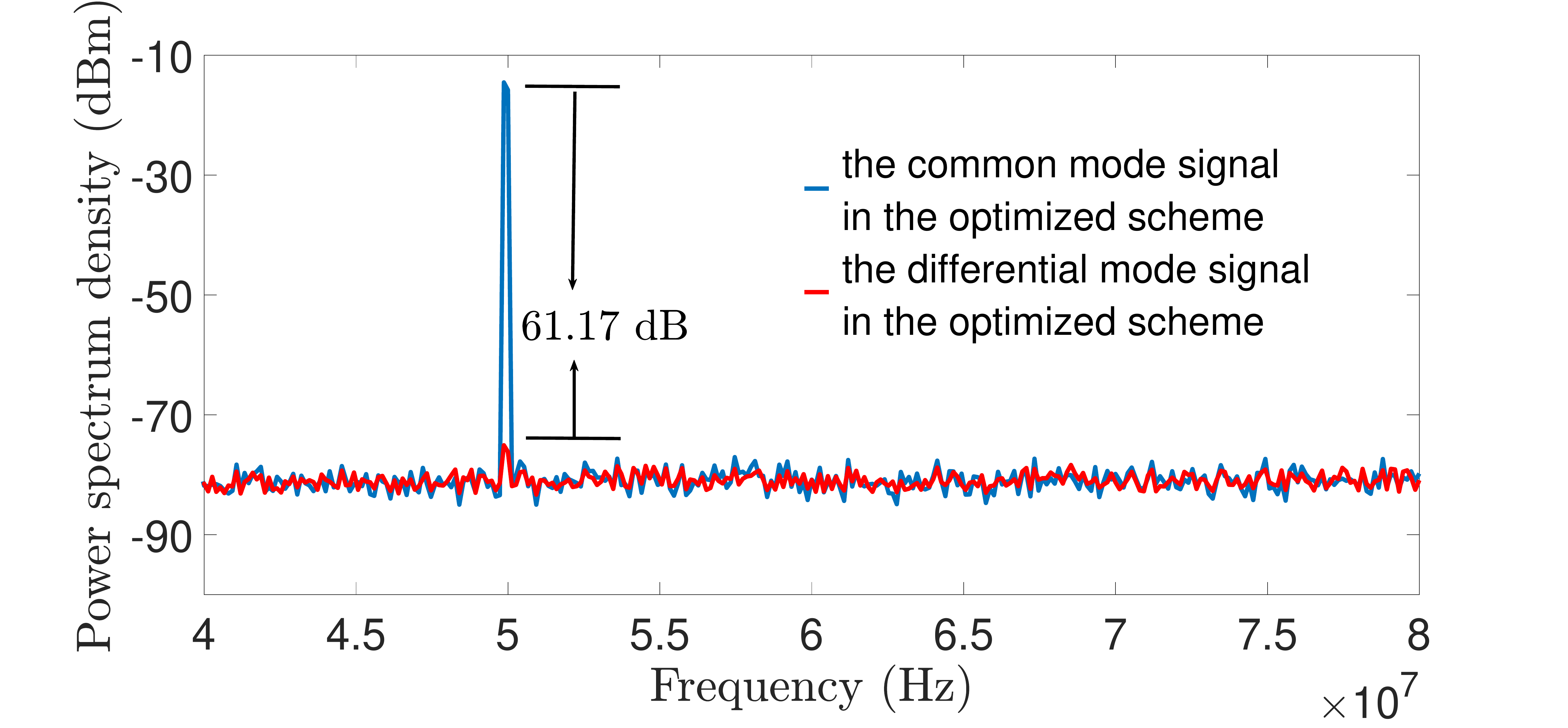}%variance_lopower4
	\caption{\label{CMRR}Power spectrums at an LO power of 40 $\mu W$ in the second routine.
		The LO is modulated by a pulsed voltage of 50 MHz. By coupling the outputs of BS into two PDs of the detector, the differential mode signal of the optimized scheme  can be obtained, as shown by the red curve. The blue curve is the common mode signal obtained by illuminating only one PD of the detector and blocking another one. The CMRR of our scheme is calculated as 61.17 dB, which indicates a significant bias elimination effect.}
\end{figure}

It should be noticed that the phase difference between two arms of the interferometer in our setup is controlled to maintain stable to support a bias-free output. However, the unbalanced MZI structure with unequal arms will be sensitive to the fluctuation of environmental temperature, which will lead to the violation of phase difference between two arms and further influence the effect of bias elimination \cite{Peng2015Broadband}. To circumvent this problem, here we have introduced several auxiliary techniques to minimize the impacts of environmental fluctuation. Firstly, we introduce an optical delay line with negligible insertion loss to make up for the length difference between two arms. In the meantime, the environmental temperature is well maintained, which makes the phase shift caused by the temperature fluctuation negligible.

Of course, there is a case where the system does not exploit the above auxiliary techniques. In this case, the fluctuated environmental temperature will cause a relatively large phase shift if the two arms of MZI are unequal, which will not only compromise the effect of bias elimination, but also cause the residual common mode noise mix into the required signal. To overcome this problem, we can also refer to the dynamic compensation method present in Ref. \cite{Zheng2019Experimental}, which realizes the compensation of phase difference in a trusted QRNG protocol. It should be noticed that it won't be a problem in integrated chips when the MZI is designed to be symmetrical.

\section{\label{set3}System performance test}

For the sake of eliminating system bias together with the common mode noise, the compensation phase ${\Delta \varphi }$ and power splitting ratio $\xi$ of the given system are adjusted as $ \pi$ and $3.6934 \times 10^4$ in the first routine and they will be set as $1.5788$ and $0.5942$ respectively in the second routine, where these values are derived based on the pre calibrated system parameters: $t_{13}$=3.7039 dB, $r_{14}$=3.7882 dB, $r_{23}$=3.7603 dB, $t_{24}$=3.7109 dB, $\eta_{PM_1}$=3.1066 dB, $\eta_{PM_2}$=3.3585 dB, $g_{PD_1}=9.93\times10^3$ V/W, $g_{PD_2}=9.69\times10^3$ V/W. These parameters help to obtain $A =19.4730$, $B =11.5700$, $C =1.8712 \times 10^3$. To quantify the capability of bias elimination, common mode rejection ratio (CMRR) is introduced to calculate the difference value between differential mode signal and common mode signal in the frequency domain.

\begin{figure}[t]
	\centering
	\includegraphics[width = 8.5 cm]{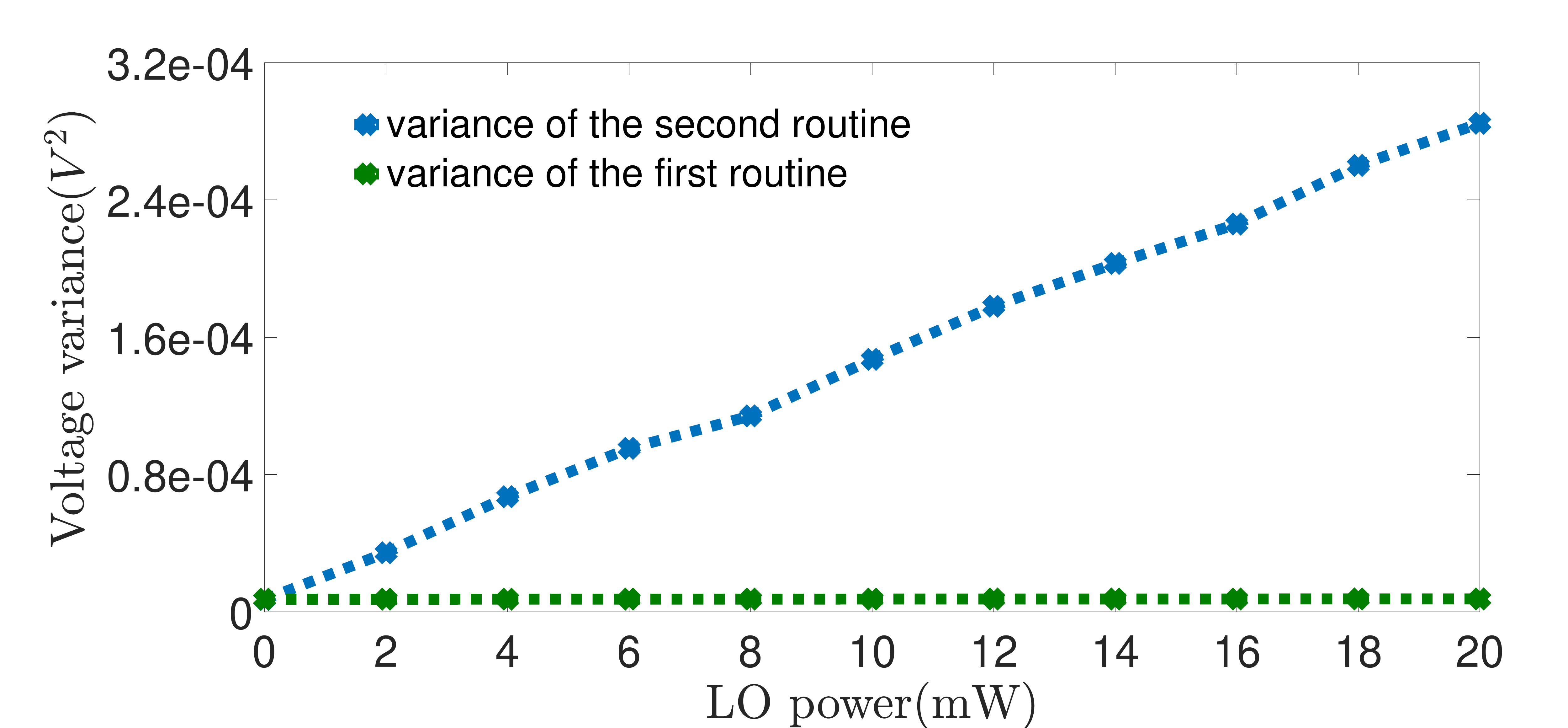}%variance_lopower4
	\caption{\label{variance_lopower}Variance vs LO power. This blue curve shows the voltage variance of the sampled raw data in the second routine as a function of the LO power, and the green curve indicates the test results of the first routine. The LO power is increased by adjusting the variable attenuator from 0 mW with a step size of 0.5 mW. In practical system, different coefficients of these two routines will lead to different slope coefficients of these two curves.}
\end{figure}

Here we take the CMRR test of the system in the second routine for an instance. The CMRR can be obtained by measuring the output spectrum of the homodyne detector and the LO applied here will be replaced by a pulsed light beam of 50 MHz with an intensity of 40 $\mu W$. As shown in Fig.\ref{CMRR}, the differential mode signal can be obtained by coupling the outputs of BS into two PDs of the detector. Simultaneously, one can record the common mode signal by illuminating only one of the PDs and blocking the other one. The CMRR can be calculated based on the maximum difference of the fundamental harmonic spectral power and the calculation result shows that the CMRR of our scheme reaches 61.17 dB, which indicates a significant effect of bias elimination. Generally, a SI-QRNG pursues not only high CMRR value, but also high bandwidth to support high random number generation rate. This is due to the reason that the bandwidth of homodyne detector will strictly limit the sampling frequency of the system to avoid large autocorrelation between sampled data, which leads to a limitation on the overall random number generation rate of a system with narrower bandwidth. Comparing with the relevant test results of the homodyne detectors with the same order of bandwidth \cite{Kumar2012Versatile,Chi2011Abalanced,Ferranti2017Anonchip,Huang2013A300MHz}, i.e., several hundreds of megahertz, our proposed scheme shows superior CMRR performance.

\begin{figure}[t]
	\centering
	\includegraphics[width = 9 cm]{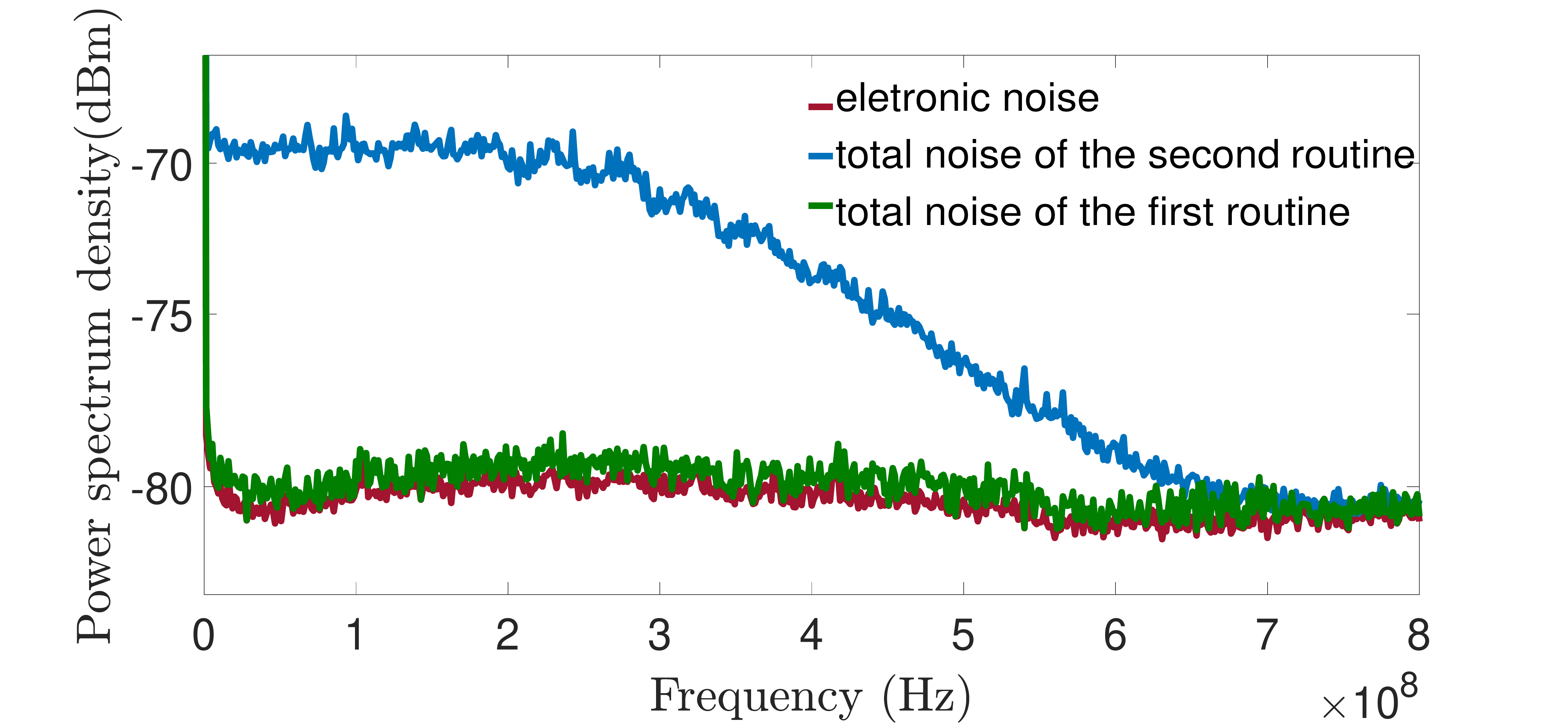}
	\caption{\label{sanli_ele} The power spectrum density of the vacuum fluctuation when the LO power is 20 mW in the first routine (green curve) and second routine (blue curve), and the electric noise when the LO power is 0 mW (red curve).}
\end{figure}

The intensity of LO should be properly set to avoid the saturation problem that causes information loss. In our experiment, the LO intensity is increased by adjusting the VOA from 0 mW with a step size of 0.5 mW and each voltage variance of measured raw data is calculated and recorded, as shown in Fig.\ref{variance_lopower}. By setting the phase of LO $\varphi  = {\pi/2}$, if the values of $\Delta \varphi $ and $\xi $ are set according to first routine, the system records the P quadrature of the input signal with a coefficient of $2{\eta _{P{M_0}}}{E_L}[A\sqrt {{t_{ac}}{r_{bc}}}  - B\sqrt {{r_{ad}}{t_{bd}}}  + C(\sqrt {{r_{bc}}{r_{ad}}}  - \sqrt {{t_{ac}}{t_{bd}}} )\cos \Delta \varphi ]$ and the test result is shown as the green curve. In the meantime, if the values of $\Delta \varphi $ and $\xi $ are set according to second routine, the system will record the X quadrature of the input signal with a coefficient of $2C{\eta _{P{M_0}}}{E_L}(\sqrt {{r_{bc}}{r_{ad}}}  + \sqrt {{t_{ac}}{t_{bd}}} )\sin \Delta \varphi $, where the test result is shown as the blue curve. In practical system, different coefficients of these two routines will lead to different slope coefficients of these two curves. We set the LO intensity as 20 mW to ensure the system works in the linear region and the power spectrum curves at the LO intensity of 0 mW and 20 mW are shown in Fig.\ref{sanli_ele}, which shows an average difference of 11.90 dB between the vacuum fluctuation at a LO intensity of 20 mW and the electronic noise within the 3 dB bandwidth, i.e., 300 MHz, in the second routine. The corresponding average difference value will be 0.88 dB in the first routine. To reduce the autocorrelation coefficients between sampled raw data, here we set the sampling frequency of ADC with 12 bit sampling precision as 600 MHz in the following experiment.

To evaluate the lower bound value of ${R_{{\rm{dis}}}}\left( {{a_{xi}}|E} \right)$, following Ref.\cite{xu2019high}, three sets data with a  length of ${n_{tot}} = 2.6214 \times {10^9}$ are obtained for evaluating the upper bound of $\overline {{V_x}} $ and $\overline {{V_p}} $, together with the Shannon entropy $H\left( {{a_{xi}}} \right)$ in three routines, where each set of data contains ${n_c} = \sqrt {{n_{tot}}}  = 5.12 \times {10^4}$ data of measured $P$ quadrature and ${n_{tot}}-{n_c}$ data of measured $X$ quadrature. The evaluated values in the first routine are $\overline {{V_x}}=2.25\times10^{-5} V^2$, $\overline {{V_p}}=2.26\times10^{-5}  V^2$ and $H\left( {{a_{xi}}} \right)=6.3274$, which corresponds to 3.3618 bit extractable random numbers. Besides, in the second routine, the evaluated results are calculated as $\overline {{V_x}}=2.85\times10^{-4}  V^2$, $\overline {{V_p}}=2.85\times10^{-4}  V^2$ and $H\left( {{a_{xi}}} \right)=8.1587$, which corresponds to 7.9107 bit extractable random numbers. The third routine measures unequal quadratures and its extractable randomness is calculated as 6.4628, where $\overline {{V_x}}=2.85\times10^{-4}  V^2$, $\overline {{V_p}}=2.25\times10^{-5}  V^2$ and $H\left( {{a_{xi}}} \right)=8.1587$. Therefore, the average extractable random bits from a single measurement in the second routine can be calculated as

\begin{equation}
	\label{Rdis2}
	\begin{split}
		{R_{{\rm{dis}}}}\left( {{a_{xi}}|E} \right) &= \frac{1}{{{n_{tot}}}}[({n_{tot}} - {n_c})(H({a_{xi}}) - S(\rho _A^G)) - t]\\
&=7.9102,
	\end{split}
\end{equation}
where $t = \left\lceil {{{\log }_2}\frac{{{n_{tot}}!}}{{{n_c}!({n_{tot}} - {n_c})!}}} \right\rceil=8.7482\times {10^5} $ is the length of random bits that control the switch of measured quadratures. The extractable randomness of the first routine is minimal in our system due to its limited signal intensity and increasing the sampling precision will help to improve its extractable randomness. Compared with the third routine with asymmetrical measured quadratures, we can find that symmetrical measured quadratures in the second routine will be beneficial to improve the extractable randomness.

\begin{figure}[t]
	\centering% for 1G-bit sequence.
	\includegraphics[width = 9 cm]{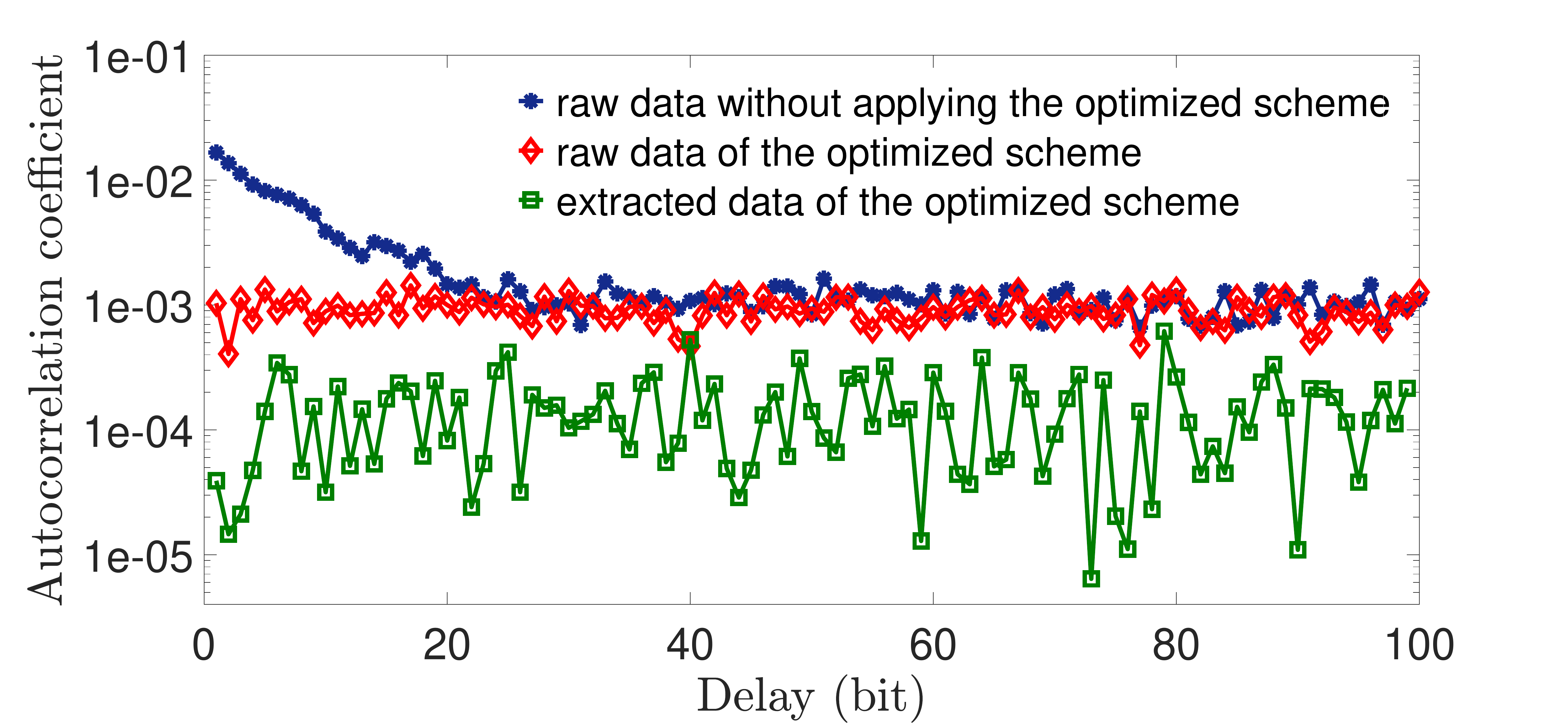}%aaa2
	\caption{\label{auto_test} Autocorrelation of 1) raw data without applying the optimized system by using the traditional scheme constructed by unbalanced devices, where the output of the $PM_0$ is directly coupled into port 1 and port 2 is blocked to provide vacuum state (blue curve), 2) raw data using the proposed bias-free scheme obtained from the second routine (red curve) and 3) its extracted data after randomness extraction (green curve). The three curves are obtained by using $10^9$ data to calculate their autocorrelation coefficients within 100 bit-delay separately.}
\end{figure}

A necessary post-processing procedure is required to the eliminate the influence of untrustworthy noise in the raw data. Toeplitz hashing function, which has the advantages of low computation and implementation complexity and provable security, is often chosen as a randomness extraction algorithm. Here, a Toeplitz hashing randomness extractor constructed by a matrix with a size of $k=3072$ columns and $j=1792$ rows is applied to eliminate the influence of untrustworthy noise in the second routine, which enables the system to reach a random number generation speed of 4.2 Gbps with a collision probability less than $\varepsilon  = {2^{ - 100}}$. Notably, the collision probability is calculated according to the leftover hash lemma $j = k \cdot R/12 - 2 \cdot {\log _2}\left( {1/\varepsilon } \right)$ \cite{Ma2012Postprocessing}.

\begin{figure}[t]
	\centering% for 1G-bit sequence.
	\includegraphics[width = 9 cm]{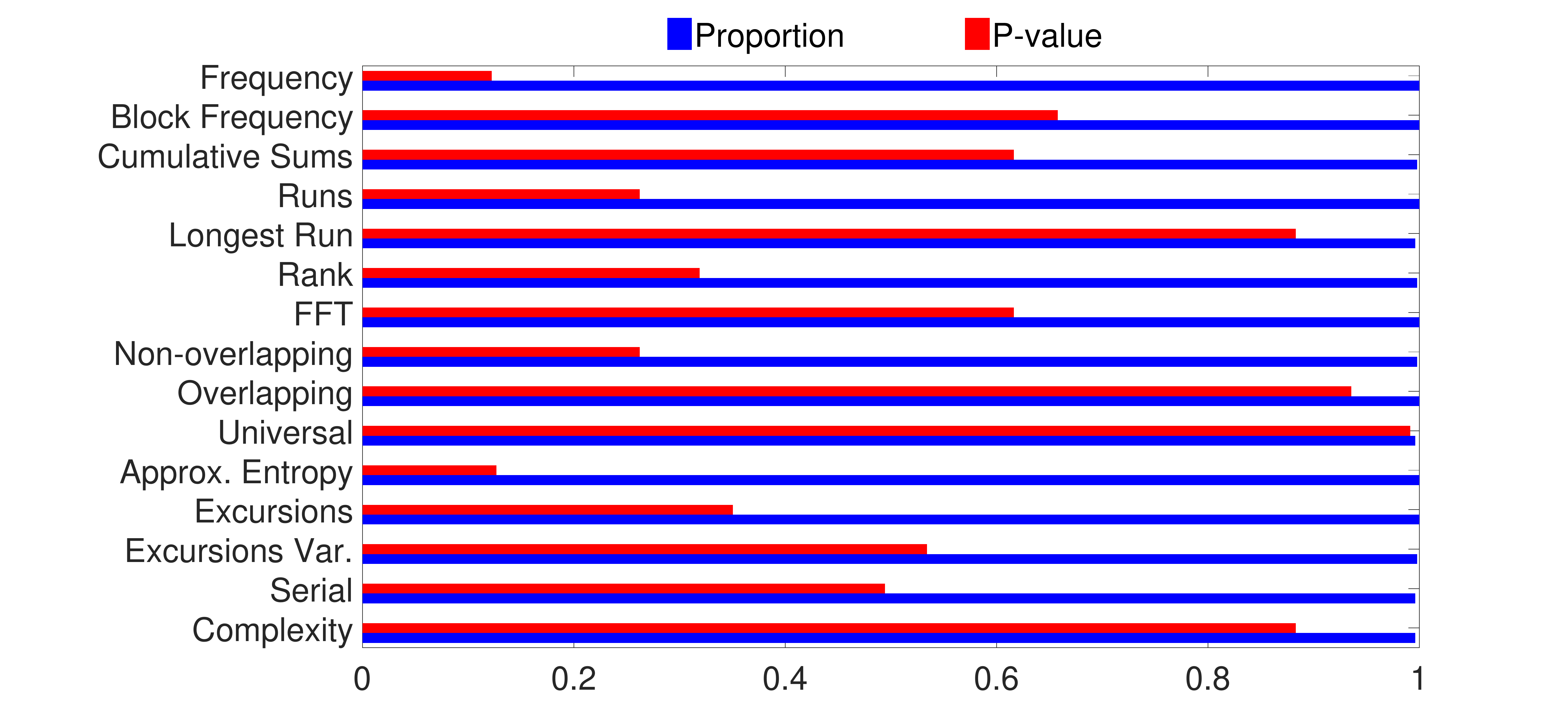}%aaa2
	\caption{\label{NIST_test_result} Results of the NIST-STS randomness tests of 1 Gbit extracted random numbers. The test suite contains 15 test items. For items with multiple P-value outputs, the Kolmogorov-Smirnov test is used to obtain a final P-value and if all P-values satisfy 0.01$ \le $P-values$ \le $0.99, the tested random numbers can be considered random.}
\end{figure}

To verify the randomness, autocorrelation tests within 100 bit-delay are calculated firstly by using three sets of data with a length of $10^9$, as shown in Fig.\ref{auto_test}.	
It should be noticed that the unbalanced system constructed by asymmetrical devices without system optimization will inevitably remain significant correlation between sampled data, which corresponds to the 2nd order autocorrelation coefficient value reaching $1.67 \times {10^{-2}}$. When applying the optimized bias elimination method by setting a proper power splitting ratio, the autocorrelation coefficients in the second routine will significantly decrease, where the average autocorrelation coefficient is reduced to $9.73 \times 10^{-4}$. Moreover, the residual correlation can be effectively reduced by applying a post-processing method, i.e., Toeplitz hashing extractor. Here, after randomness extraction, the coefficients will be less than $1.50 \times 10^{-4}$, which indicates the correlation between these extracted random numbers is not significant. Then we apply the NIST-STS suite for randomness test, and the test results are shown in Fig.\ref{NIST_test_result}, which indicate that the random bits generated by the proposed SI-QRNG scheme can pass all the test items.
%%%%%%%%%%%%%%%%%%%%%%%%%%%%%%%%%%%%%%%%%%
\section{\label{set4}Conclusion}

In this paper, we have proposed and experimentally demonstrated an optimized bias-free SI-QRNG scheme by exploiting an all-optical method for the elimination of system bias and common mode noise introduced by the fluctuated LO. The scheme explores a bias-free SI-QRNG structure suitable for system integration based on the existing technologies. Besides, the system parameters are optimized to seek for measuring only a single quadrature in each measurement, which can effectively circumvent the problem of simultaneous measuring two quadratures in a single measurement and can be further exploited to realize the SI-QRNG system under three different routines. Particularly, by assuming that the source is untrustworthy, we set the system to the second routine and randomly switch the phase of $PM_0$ to realize the measurement of two quadratures, which enables the system to support up to 4.2 Gbps source-independent random number generation. Compared with the third routine, it is verified that the symmetrical measurement of two quadratures is beneficial to obtain a faster random number generation speed than the asymmetrical case.

Notably, the MZI structure presented in our manuscript serves the purpose of a reconfigurable beam splitter in the experiment and it can be replaced by a simplified scheme based on a tailored fixed beam splitter and a single-phase modulator, where an additional optic fiber patch cable will be required to make up for the length difference between the two arms due to the reason that its length difference exceeds the compensation range of the optical delay line. What's more, the proposed SI-QRNG scheme could be easily integrated into the silicon photonic chip for continuous-variable quantum key distribution system \cite{zhang2019integrated}. It will make continuous-variable quantum key distribution system \cite{zhang2019continuous,zhang2020long} low cost and high practical security in the future. It should be noticed that the proposed system is named as SI-QRNG to unify with previous works\cite{marangon2017source,xu2019high}. The expression of semi-source-independent QRNG will be more accurate because the system still assumes an i.i.d input \cite{Drahi2019Certified}. For further study, applying the proposed protocol into the integrated continuous-variable quantum key distribution and evaluating its performance in practical integrated chips will be interesting. Simultaneously, building a theoretical model for the effects of local oscillator intensity fluctuations in the SI-QRNG scenario and quantifying extractable randomness through quantum analysis \cite{zhou2018randomness} will be valuable work and we would like to include them in future work.

\section*{Funding}
National Natural Science Foundation of China (61531003, 61427813); Fund of CETC (6141B08231115); Fund of State Key Laboratory of Information Photonics and Optical Communications.

\end{document}